\documentclass{amsart}
\usepackage{graphicx,euscript}
\theoremstyle{plain}
\newtheorem{Thm}{Theorem}[section]
\newtheorem{Cor}[Thm]{Corollary}

\theoremstyle{definition}

\newcommand{\wt}{\widetilde}

\newcommand{\lp}{\bigl(}
\newcommand{\rp}{\bigr)}

\newcommand{\gs}{\sigma}

\newcommand{\gd}{\delta}
\newcommand{\gl}{\lambda}

\newcommand{\gep}{\epsilon}

\newcommand{\nc}{\newcommand}

\newcommand{\eu}{\EuScript}
\newcommand{\indic}{\boldsymbol{1}}

\newcommand{\on}{\operatorname}
\nc{\G}{\eu{G}}
\nc{\lip}{\on{Lip}}
\nc{\izf}{\int_0^\infty}
\nc{\tand}{\text{ and }}
\nc{\tst}{\text{ s.t. }}
\nc{\fM}{\mathfrak{M}}
\nc{\fP}{\mathfrak{P}}

\newcommand{\lb}{\bigl[}
\newcommand{\rb}{\bigr]}
\newcommand{\lv}{\bigl|}
\newcommand{\rv}{\bigr|}

\newcommand{\R}{\mathbb{R}}

\nc{\N}{\mathbb{N}}
\nc{\mL}{\eu{L}}
\nc{\mA}{\eu{A}}
\nc{\mM}{\eu{M}}
\nc{\C}{\eu{C}}
\nc{\B}{\eu{B}}

\newcommand{\dbl}{\renewcommand{\baselinestretch}{1.3}\normalsize}

\nc{\vx}{\vec{x}}
\nc{\vy}{\vec{y}}
\nc{\DF}{\eu{F}}
\nc{\df}{f}
\nc{\tX}{\wt{X}}
\nc{\mE}{\mathbb{E}}
\nc{\brM}{\bar{\mM}}
\nc{\tih}{\tilde{h}}
\nc{\lep}{\frac{\gl}{\gep}}
\nc{\tp}{(2\pi)^{-1/2}}
\nc{\sM}{\mM^{*}}
\nc{\ns}{\nu^*}
\nc{\bP}{\bar{P}}
\nc{\bS}{\bar{S}}

\begin{document}
\title[Mutation-selection model]{
A generalized model of
mutation-selection balance with applications to aging}
\author{
David Steinsaltz$^{+}$,  Steven N. Evans, and Kenneth W. Wachter}
\begin{center}
\end{center}
\begin{abstract}
A probability model is presented for the dynamics of
mutation-selection balance in a haploid infinite-population
infinite-sites setting sufficiently general to cover
mutation-driven changes in full age-specific demographic schedules.
The model accommodates epistatic as well as additive
selective costs.   Closed form characterizations are
obtained for solutions in finite time, along with
proofs of convergence to stationary distributions and
a proof of the uniqueness of solutions in a restricted case.
Examples are given of applications to the biodemography of
aging.
\end{abstract}
\renewcommand{\thefootnote}{}
\maketitle

\dbl
\section{
Introduction}
\label{sec:intro}
\footnote{$^{+}$
To whom correspondence should be addressed at Department of Demography,
University of California, 2232 Piedmont Avenue, Berkeley, CA. 94720-2120;
email dstein@demog.berkeley.edu.}

Arguments from the mathematical genetics of
mutation-selection balance figure broadly in evolutionary theories
of senescence.  Available formal models, however,  do not cover cases
brought to the fore by recent progress in biodemography \cite{kW03}.
In this paper,  we present a rigorous general model encompassing 
these cases,   prove results concerning existence, uniqueness,
and convergence,  obtain closed-form representations for solutions
to the model,  and give examples of its application to questions
in the demography of aging.

The whole mathematical theory of natural selection may be
divided into three parts:
positive mutations, neutral mutations, and deleterious mutations.
Positive mutations may be thought to add up to an optimal adaptation,
at least under some conditions,
and they are generally studied in that context by demographers. 
Neutral mutations have their primary effects in alleles which 
drift randomly to fixation.
Deleterious mutations, the focal subject for theories of aging
and for this paper, are expected never to achieve fixation in
populations, except, through founder effects,  in very small populations.
Their influence in large populations derives from their
persistent reintroduction and slow meander to extinction.

Sir Peter Medawar \cite{pM52},
in 1952, descried an explanation for senescence in the
accumulation of deleterious alleles with age-specific effects,
given the declining force of natural selection with adult age.
W. D. Hamilton \cite{wH66}
presented expressions for this declining age-specific force, helping
others quantify the resulting balance between mutation and selection.
B. Charlesworth \cite{bC94}
analyzed the dynamics of age-specific selection. His work guides
the thinking of many experimentalists.  

At stake are the cumulative effects of numerous mildly deleterious
mutations showing up at some large collection of loci.
In our setting, the genotypes determine full age-specific
schedules of mortality and fertility,  and the effects of a
mutation have to be represented as a perturbation of a whole
function of age.  A rigorous treatment demands that mutations
correspond to points in abstract spaces, such as function spaces.
Relationships between our work and the large literature on
mutation and selection reviewed by B\"{u}rger \cite{rB00} 
are discussed in Section \ref{sec:prior}.

Up to now,  researchers have relied on linear 
approximations to cost functions and restricted their 
representations of the age-specific effects of mutations to 
stylized patterns like step-functions.  Intriguing results have 
been obtained.  Some are discussed in Section \ref{sec:longevity}.    
The linear analysis, however, can be deceptive, 
and the stylized patterns are remote from realistic portrayals
of gene action.  Cases chosen for analytic tractability  
give a misleading picture of the full range of possibilities.

Our model is an infinite-population, multiple-sites or
infinite-sites  model in continuous time.
The dynamical equation is a fairly standard one,  but the
space of mathematical objects to which it applies is novel.
Our model allows a highly flexible specification of pleiotropic
gene action.   It is especially suited to demographic applications
with mutant alleles affecting age-specific schedules.
The model is a haploid infinite-population model with no 
recombination.   A parallel model with free recombination,
introduced in Section \ref{sec:prospects}, will be developed 
in a future paper.

Our contribution is to allow large numbers of interacting
genes to make small contributions to a continuum of linked traits.
Traditional analyses which recognize individual alleles
(thus admitting, in principle, arbitrary configurations of pleiotropy)
are amenable only to small numbers of loci;
quantitative genetics, which reduces the contributions of
individual genes to a continuum, reduces the complexity of
pleiotropy to covariance matrices.

Although multi-locus models without recombination like our own 
can be formally imbedded in single-locus models,  this imbedding
will not generally yield useful results.  When a multilocus model is
translated into the single-locus framework, it brings along an
extra structure of transition rates, whose complexity grows 
exponentially with the number of loci.  When the number of 
loci is large or, as in our model, effectively infinite,
this extra structure overwhelms the single-locus infrastructure.
In our function-space setting,  the formal embedding itself
also poses difficulties.  As a consequence, results for
single-locus models are mainly helpful as analogies.   

Unlike most models of which we are aware,  our model comfortably
accommodates epistasis.  
(A very different approach to epistasis, in the two-allele 
setting, may be found in \cite{ZP98}.)  
The selective cost of a mutant allele can
depend on the configuration of other mutant alleles present in a genome.
This property is critical to the study of senescence, even
without special assumptions about interactions among genes, because
the fitness costs of cumulative demographic changes are not linear.

We are able to obtain closed-form representations of the entire
time path of solutions to our dynamical equation (Theorem \ref{T:epistatic}).  Our results
are not restricted, like much previous work,  to limiting states
and equilibrium distributions.   We give proofs of convergence 
over time (Theorem \ref{T:epiform}),  and set machinery into place
to compute rates of convergence  and to cope with changing fitness
conditions as well.  In Section \ref{sec:asymptotic}, we present 
some results about the asymptotic behavior of solutions.  
Theorem \ref{T:blowup} gives sufficient conditions for the 
numbers of certain classes of mutant alleles to increase 
without limit, generalizing the well-known ``error threshold'' 
(cf., \cite{WK93}.)  In Section \ref{sec:nonepistatic} 
we derive the Poisson limit for the non-epistatic case, 
as well as proving uniqueness of the solution.
In the general epistatic case we do not yet have a proof of uniqueness.  
In Section \ref{sec:longevity} we discuss some implications 
of our results for the theory of longevity.  In Section \ref{sec:prior} 
we review earlier work on related problems.

\section{
The model}
\label{sec:model}
We consider an infinite population subject to mutation and selection.
There is a complete, separable metric space $\eu{M}$ of potential
mutations, on which is defined a boundedly finite Borel measure $\nu$.
(In other words, $\nu$ assigns finite mass to bounded sets; together with the
assumptions on $\eu{M}$, this condition implies that $\nu$ is $\sigma$-finite.)
We refer to this measure as the ``mutation rate''; for any set $B$, the
quantity $\nu(B)$ represents the rate at which there spontaneously
arises a mutant allele from $B$.  Our picture is one in which
new mutant alleles are steadily arising, each one tagged by a
corresponding point of $\mM$.  For convenience, we identify the tag
with a description of the effects that the mutant allele produces:
for instance, a function on the non-negative real line $\R^{+}$ 
giving the increases in mortality attributed to the action 
of that allele at each age.

The space of ``genotypes'' $\eu{G}$ is identified with the
integer--valued boundedly finite Borel measures on $\eu{M}$,
with a topology to be described shortly.
An element of $\eu{G}$ has the form $\sum \delta_{m_{i}}$,
where the $m_{i} \in \eu{M}$  are not necessarily distinct and
the number of $m_{i}$ in any bounded subset of $\mM$ is finite.
The notation $\delta_x$ stands in general for a unit mass
at the point $x$ in the space to which $x$ belongs.
Each genotype represents a set of mutant alleles
that an individual may carry.  The ``null genotype''
has wild-type alleles at every locus and carries none of these
mutant alleles.

The state of the population at time $t$ is denoted $P_{t}$,
which is a Borel probability measure on the measures in
$\eu{G}$.  Thus $ P_{t}$ is the distribution of a random
measure \cite{oK02, DV88}.
The evolution of the population is presumed to be so slow
that it can be represented as occurring in continuous time,
without reference to discrete generations.

To each genotype $g$ we assign a``selection cost'' $S(g)$;
$S$ is a continuous function from $\eu{G}$ to $\R^+$.
(Including negative costs would be feasible for the finite-time solutions,
at the expense of slightly more complicated statements for theorems.)
In applications, costs will typically be decrements
to growth rates, in effect measuring fitness on a logarithmic scale.

We normalize costs so that $S$ vanishes on the null genotype,
and vanishes for no other $g$.
On $\eu{M}$ we write $S(m)$ for the cost of the
singleton $ g = \delta_m$.   When $S$ is linear, so that
$S(g+\delta_{m})-S(g)$ is independent of $g$,
the model is additive, or {\em nonepistatic}.

Any measure $P$ on $\eu{G}$, like $P_t$,
may be determined by the expectation values it assigns
to a suitably rich collection of functions
$F$ from $\eu{G}$ to $\R$, as specified below.
For brevity,  we write  $PF$ or $P(F)$ for the
expectation value $ PF = \int_{\G} F(g) dP(g) $
of any measurable function from $\G$ to $\R$ such that
$ \int |F(g)| dP(g) < \infty$.
Since genotypes are measures, we can also write
$  gf = g(f) = \int f(m)dg(m)=\sum f(m_{i}) $
when $f:\mM\to\R$, and $g=\sum \delta_{m_{i}}$.

Our dynamic equation for $P_t$ is
\begin{equation}  \label{E:evfunc}
   \frac{d}{dt}P_{t}F
      = P_{t} \left( \int \lb F(\cdot+\delta_{m}) -
      F(\cdot) \rb d\nu(m) \right) \, - \, P_{t}(FS) \, + \,
      (P_{t}F)(P_{t}S)
\end{equation}

The meaning of the equation is readily described when $F$ is
the indicator function of a set $G$ of genotypes.  The first
term inside the integral measures the rate at which the
population is flowing into the states in $G$ out of all sorts
of other states because of the addition of a mutant $m$ that
lines up just right to enter $G$.  The second term
inside the integral measures the rate at which population
flows out of $G$ because of new mutations.   The remaining
two terms measure the effect of selection.  The proportional
rate of change in mass of the population in $G$ equals
the difference between the average fitness cost of genotypes in
$G$ and the average fitness cost of the whole population.

Measuring fitness relative to the changing average fitness
of the whole population keeps total mass constant and lets
the measure $P_t$ represent the probability of finding a
randomly selected individual in a given state,
modeling population distribution rather than population size.
While our equation may be novel, it is strongly
analogous to standard mutation-selection dynamics on quantitative
traits, such as those given in equation V.2.11 of \cite{rB00}.

When mutations are identical (so that $\eu{M}$ comprises
only a single point)  we have the ``mutation-counting model''
going back to Kimura and Maruyama \cite{KM66},
whose history will be described in Section \ref{sec:prior}.
A genotype is specified by a natural number, the number
of mutant alleles present in it, and \eqref{E:evfunc}
becomes
\begin{equation}  \label{E:timeevolsimp}
     \frac{ dP_t(n) }{dt}  =   \nu P_t(n-1) - \nu P_t(n)
      -  P_t(n) \left( S(n) - \sum_m S(m) P_t(m) \right )
\end{equation}
In the non-epistatic case, where $S$ is additive,
the mutation-counting model (or its discrete-time counterpart)
has a Poisson distribution with parameter $\nu/S(1)$
as its stationary distribution.

For general $\eu{G}$,  the counterpart of the Poisson distribution
is a Poisson random measure.  For the non-epistatic case,
Theorem \ref{T:additive} establishes conditions for uniqueness
and convergence to a stationary distribution given by a Poisson
random measure with intensity $ (1/S(m)) d \nu(m) $,
the measure on $\eu{M}$ whose Radon-Nikodym derivative with
respect to $\nu$ is $ 1/S $.  (A Poisson random measure assigns 
a Poisson-distributed random integer mass to each measurable set,
the mean of the mass assigned to a set is the intensity measure
of the set, and the random masses of disjoint sets are
independent random variables.) The general 
theory \cite[Chapter 7]{DV88} takes care of technical details.
Even in the non-epistatic case, only rather special starting 
states lead to the Poisson limit.
In the epistatic case,  covered by Theorem \ref{T:epistatic},
asymptotic distributions, when they exist, may not be Poisson.

We need a suitable notion of weak convergence
for boundedly finite random measures.  Following Appendix A.2
of \cite{DV88}, we equip the space $\G$ with the metrizable
$\hat w$-topology under which a sequence of measures
$g_1, g_2, \ldots \in \G$ converges to a measure $g \in \G$
if and only if $\lim_n g_n(f) = g(f)$ for each bounded
continuous function $f:\mM \rightarrow \R$ that is supported
on a bounded set.  A sequence of probability measures
$Q_1, Q_2, \ldots$ on $\G$ (that is, a sequence of distributions
of boundedly finite random measures on $\mM$) converges weakly
to the probability measure $Q$ on $\G$ with respect to the
$\hat w$-topology if and only if $\lim_n Q_n(F) = Q(F)$
for every bounded $\hat w$-continuous funtion $F$ on $\G$.
This turns out to be equivalent to the requirement
that $\lim_n Q_n(F) = Q(F)$ for all $F$ of the
form $F(g) = e^{-g(f)}$ for some continuous boundedly
supported $f:\mM\to\R^{+}$.  This class $\eu{F}$ is the sufficiently
rich class of functions required for our expectation
values to determine our measures:  Equality of expectations for
$F$ in $\eu{F}$ implies equality of expectations for
all bounded Borel-measurable $F$ \cite[Section 6.4]{DV88}.

We must bear in mind that ordinary theorems guaranteeing existence
and uniqueness of solutions to differential equations do not extend
to our abstract setting.  The derivatives on the left-hand side
of \eqref{E:evfunc} might not exist, and, in the presence
of unbounded $S$, we might have infinity minus infinity
on the right-hand side.  Our proofs are constructive,
so the meaningfulness of the equation will follow from the
properties of the proferred solutions.  We prove a reasonable
version of uniqueness in the non-epistatic case.  In the general
epistatic case, we have not yet ruled out multiple alternative
meaningful solutions; there could be a complicated mathematical
question lurking here.

\section{
Existence of solutions}
\label{sec:main}
We express the solution in terms of a certain random meaure on
${\mathcal M} \times {\mathbb R}^+$.  We let $\Pi$ denote the
Poisson random measure on this space with intensity measure
$\nu \otimes {\mathrm{Lebesgue}}$.
Define a time-homogeneous $\eu{G}$-valued Markov process
$ (X_t)_{t \ge 0 } $
by
\begin{equation}
X_t := X_{0}+\int_{ \eu{M} \times [0,t]} \delta_m \, d\Pi(m,u),
\end{equation}
where $X_{0}$ is a random measure with distribution
$P_{0}$, independent of $\Pi$.
Each realization of $X_{t}$ may be pictured as a discrete set of points,
possibly with duplication; as time passes, new points accrete.
The cost function $S$ could be allowed to depend on time,
but we keep time-independent notation here.  

\begin{Thm}  \label{T:epistatic}
Suppose that there is a positive $T$ such that
\begin{equation} \label{E:defTep}
\mathbb{E} \exp \left( - \int_{0}^{t} S(X_{u})du
     \right) S(X_{t}) < \infty
 \end{equation}
for all $t\in [0,T)$. Then the equations \eqref{E:evfunc} 
have a solution on $[0,T)$, given by
\begin{equation}  \label{E:epissol}
    P_t F = \frac
   { {\mathbb E} \left[\exp\left( - \int_0^t S(X_u) \, du\right)
    F(X_t)\right] }
   { {\mathbb E} \left[\exp\left( - \int_0^t S(X_u) \, du\right)
      \right]}.
\end{equation}
\end{Thm}

\begin{proof}
Define a linear operator on the continuous functions
on the genotype space by
\begin{equation}
  A F = \int \left[ F(\cdot+\delta_{m}) - F(\cdot) \right] d\nu(m)
      - S(\cdot) F(\cdot) .
\end{equation}
Given an integrable function $\sigma(t)$,
put $\tilde P_t = \exp\left(-\int_0^t \sigma(u) \, du\right) P_t$.
If we can arrange for $ \sigma(t) $ to equal the average
selective cost $ P_t S $, then, thanks to the chain rule,
the derivative of $ \tilde P_t F $ must equal $ \tilde P_{t} AF $.

The operator $A$ may be unbounded if $\nu$ has infinite total mass,
but it is well-defined on the class $\eu{F}$ and
is the generator of a sub-Markovian semigroup
$(\Gamma_t)_{t \ge 0}$.  By the Feynman-Kac formula
\cite[Section III.19]{RW00}, $\Gamma_{t}$ may be described as
\begin{equation}
 \Gamma_t F(g) = {\mathbb E} \left[\exp\left( - \int_0^t S(g+X_u-X_{0}) \,
 du\right) F(g+X_t-X_{0}) \right].
\end{equation}

Now, the semigroup $(\Gamma_t)_{t \ge 0}$ solves the forward equation
$  \frac{d}{dt} \Gamma_t F = \Gamma_t (A F) $
and it follows that
$ \tilde P_t F = \tilde P_0 \Gamma_t F $, which equals the
numerator of \eqref{E:epissol}.  By the condition on $T$,
$ \tilde P_t S $ is finite on $[0,T) $, equalling the derivative
of $ \tilde P_t I $,  so that we may
put $ P_t F =  \tilde P_t F / \tilde P_t I $ and achieve
$ \sigma(t) = P_t S < \infty $ on [0, T).
\end{proof}

\section{
Representations}
\label{sec:represent}
Although our solution \eqref{E:epissol} may look abstract,
as long as $\nu(\eu{M})$ is finite $P_t F $ can be expressed  
as a series expansion whose terms can be evaluated by
multiple integration.  We now derive this 
expansion, which makes direct calculations feasible
in applications.   
When $\nu(\eu{M}) $ is finite, we order the points
put down by $\Pi$ according to their arrival
times $ \tau(1), \tau(2) \ldots $ and
write $ Y_n := X_{\tau(n)} $.   Let $J_n$ be
the indicator function of genotypes with exactly $n$
(possibly overlapping) points: $ J_n(g) = 1 $
if  $g(\eu{M})=n$, and 0 otherwise.
Our series expansion for $ P_t F $ will take the
form  $ \sum_n P_t J_n F $. 
We write $ x \wedge y $ for the lesser of any two
quantities $x$ and $y$.  Renewal theory calculations
turn \eqref{E:epissol}  into a handy formula
for probabilities of $n$-point genotypes:

\begin{Thm}  \label{T:epiform}
Suppose $ \nu(\eu{M}) < \infty $ and $P_0$ puts
unit mass at the null state $0$, with $S(0)=0$.
Then the solution \eqref{E:epissol} may be written as
$ P_t F =  \tilde P_t F / \tilde P_t \indic $,
with
\begin{equation}\label{E:ptn}
 \tilde P_t J_n F =  \nu(\mM)^{n}e^{-\nu(\eu{M})t}
      \mE \left[\lp S(Y_1) \ldots S(Y_n) \rp^{-1}  H_{t,n} F(Y_n)\right].
\end{equation}
Here $H_{t,n}$ is a conditional probability defined in terms
of independent unit-rate exponential variables $Z_1, Z_2, \ldots $
by the formula
\begin{equation}\label{E:expo}
 H_{t,n} = \mathbb{P}
        \left\{ \sum Z_{j}/S(Y_{j})<t \, | \, Y_1, \ldots Y_n \right\}
\end{equation}

If $ \sum \nu(\mM)^{n}\mE  [( (S(Y_1) \ldots S(Y_n) )^{-1}] $ is finite,
$P_t$ converges in distribution as $t$ goes to infinity.
If the sum is infinite,  $ P_t J_n $ goes to zero for all $n$.
\end{Thm}

\begin{proof}
Consider the numerator of \eqref{E:epissol} with $J_nF $ in
place of $F$.     The integral inside the exponential is the sum of
terms $ S(Y_j)( \tau(j+1) \wedge t - \tau(j))$ for $j$ from $1$
to $n$.
The factor $J_n$ restricts the domain to
the event $ \{ X_t(\eu{M}) = n \}$, an event with
probability $ e^{-\nu t} \nu^n t^n/n! $, where we write
$ \nu $ for $ \nu(\eu{M}) $. Conditional on this event,
the $Y$'s are independent of the $\tau$'s,  and the
$\tau$'s are distributed like the order statistics
of a sample of $n$ uniform random variables
$u_1 \ldots u_n $ on $[0,t]$ which may occur in any
of $n!$ orderings. Put $u_{n+1} = t$.

To obtain the expectation over the $ \tau$'s,
we evaluate the integral
\begin{equation}
n! t^{-n}    \int_{0}^{t} \int_{u_{1}}^{t}\cdots \int_{u_{n-1}}^{t}
  \exp\left\{-\sum_{i=1}^{n} S(Y_i)(u_{i+1}-u_{i})\right\}
   du_{n}\cdots du_{2}du_{1},
\end{equation}
The change of variables  $ z_i = u_{i+1} - u_{i} $ transforms
this integral into the product $ n! /( t^n S(Y_1)\ldots S(Y_n)) $
times
\begin{equation}
\int\cdots\int
     \left(S(Y_1)e^{-S(Y_1)z_{1}}\right)\ldots 
     \left(S(Y_n)e^{-S(Y_n)z_{n}}\right)
     dz_{1}dz_{2}\ldots dz_{n},
\end{equation}
The integrations range over all non-negative $z_1 \ldots z_n$
such that  $z_{1}+\ldots +z_{n}< t $, yielding the
exponential probability expression $ H_{t,n}$.
Closed-form formulas for $H$ are given in \cite[Ch. 1, 13.12]{wF71}.
The probability $ e^{-\nu} \nu^n t^n/n! $ times
$ n! /(t^n S(Y_1)\ldots S(Y_n)) $ times $ H_{t,n} $
gives \eqref{E:ptn}.

We bound $ \tilde P_t J_n S $
by $ \nu \tilde P_t J_{n-1} $, noting
that $ H_{t,n} \le H_{t,n-1} $.  Summing over $n$, we
find $ \tilde P_t S \le \nu \tilde P_t \indic \le \nu $,
verifying the supremum condition
for all finite $T$.
The factors of $ e^{-\nu t}$  in the numerator
and denominator of $P_t J_n F$ cancel.  The conditional
probability $ H_{t,n}$,  is monotone increasing in $t$
toward a limit of $1$ for each choice of $n$
and $Y_{1},\dots,Y_{n}$. Hence the limit claim follows
by monotone convergence.
\end{proof}

In demographic applications we are typically interested
in counting the average number of mutant alleles of a 
given type that a randomly chosen individual would bear.
For $B$ a measurable subset of $\eu{M}$,
write $R_{t}(B)$ for the expected number of mutations from $B$
at time $t$; that is, $R_{t}(B)=\int_{\G} g(B) dP_{t}(g)$.
For special starting states, we can obtain a closed-form density
for $R_t$.

\begin{Thm}  \label{T:palm}
Suppose the starting distribution $ P_0 $ is a Poisson measure
with intensity $\pi_{0}$.   Then the measure $R_{t}$ has
the form $ \zeta_t(m) d\nu(m) + \eta_t(m)d\pi_0(m) $ where
\begin{equation}  \label{E:palm}
\begin{split}
\zeta_t(m) &= \frac{ \mE  \lb\exp\bigl(-\int_{0}^{t}S(X_{u})du\bigr)
        \int_{0}^{t}\exp\bigl(-\int_{\tau}^{t}
          [S(X_{u}+\delta_{m})-S(X_{u})]du\bigr) d\tau ]}
       {\mE [\exp\bigl(-\int_{0}^{t}S(X_{u})du\bigr)]}          \\
\eta_t(m) &=
 \frac{\mE\lb\exp\bigl(-\int_{0}^{t}S(X_{u}+\gd_{m})du\bigr)\rb}
       {\mE\lb\exp\bigl(-\int_{0}^{t}S(X_{u})du\bigr)\rb }.
\end{split}
\end{equation}
\end{Thm}
\begin{proof}
When the initial distribution is Poisson,  the entire process $X_t$,
including $X_0$, is defined from a Poisson random
measure $ \xi = \Pi + (X_0, \delta_0) $ on the product space
$ \eu{M} \times \mathbb{R}^+ $ with intensity measure
$H = \nu \otimes {\mathrm{Lebesgue}} + \pi_0 \otimes \delta_0 $.
The local Palm distribution
for the Poisson random measure $\xi$ at $ (m, \tau) $
in  $ \eu{M} \times \mathbb{R}^+ $ is the distribution
of $\xi$ itself augmented by an atom
at $ (m, \tau) $ \cite[Example 12.1(b)]{DV88}.
For any non-negative bounded Borel-measurable function $G(m, \tau, \xi)$
the Palm integral formula \cite[Proposition 12.1.IV]{DV88} makes
\begin{equation}\label{E:palminteg}
    \mathbb{E} \int G(m,\tau, \xi)  d\xi(m,\tau)
    = \int \mathbb{E} \, G(m,\tau, \xi+\delta_{(m,\tau)}) dH(m,\tau)
\end{equation}
The integrals are taken over $ \eu{M} \times \mathbb{R}^+ $,
and $ \mathbb{E}  $ operates on $\xi$.
Fix $t$ and $ B \subset \eu{M} $ and choose the function $G$ to
be
\begin{equation}\label{E:Gmtauxi}
G(m,\tau, \xi)=  \exp\left( - \int_0^t S(X_u) \, du \right)
           \indic_{\{m\in B\}}  \indic_{\{\tau \le t\}}.
\end{equation}
Bear in mind that $X_u$ is a function of $\xi$.
With this $G$, plugging into Equation \eqref{E:epissol},
$ \tilde{P}_t X_t(B) $ is given by the left-hand side
of \eqref{E:palminteg}.   On the right-hand side, the
extra atom at $(m,\tau)$ changes the argument of the exponential
function inside \eqref{E:Gmtauxi}
into $  - \int_{0}^{\tau} S(X_u)du - \int_{\tau}^{t} S(X_u + \delta_m ) du$.
The first term in $H$, which is $ \nu \otimes {\mathrm{Lebesgue}} $, calls
for integration over $\tau$ and gives the contribution in the
numerator of $\zeta$ in \eqref{E:palm} with respect to $\nu$.
The second term in $H$ puts $\tau $ equal to zero and gives the
contribution in the numerator of $\eta$ with respect to $\pi_0$.
The denominator in $\zeta$ and $\eta$ is a constant independent
of the set $B$. It converts $\tilde{P}_t$ to $P_t$.
The indicator function in $G$ arranges that the measure $R_t(B)$
is obtained by integrating over $B$, so $\zeta$ and $\eta$ are
indeed Radon-Nikodym derivatives for $R_t$ as claimed.
\end{proof}

When the process $P_t$ starts from the null genotype
we set $\pi_0 = 0 $.
Equations \eqref{E:palm} allow us to compare the influences
of different cost functions:

\begin{Cor}  \label{C:compare}
Assume that the conditions of Theorems \ref{T:epistatic} and
\ref{T:palm} are satisfied, and suppose
that $S$ is sub-additive; that is, $S(g+g')\le S(g)+S(g')$.
Define the corresponding additive cost
function $\bar{S}(g):=\int S(\delta_{m}) dg(m).$
Let $P_{t}$ and $\bar{P}_{t}$ be corresponding genotype
distributions produced by \eqref{E:epissol}.
Then $P_{t}F\ge \bar{P}_{t}F$ for any linear $F$ of the
form $F(g)=g(f)$, where $f$ is nonnegative, measurable,
and has bounded support.
\end{Cor}

\begin{proof}
The sublinearity of $S$ and the linearity of $\bar{S}$ imply
\begin{equation} \label{E:sublin}
\begin{split}
 \zeta_{t}(m) &\ge
   \frac{\mE\lb  \exp\bigl(-\int_{0}^{t}S(X_{u})du\bigr)
       \int \exp\bigl(-(t-\tau)S(\delta_{m})\bigr) d\tau \rb }
        {\mE \exp\bigl(-\int_{0}^{t}S(X_{u})du\bigr)} \\
       &=  \frac{1-e^{-S(m)t}}{S(m)}
 = \bar{\zeta}_{t}(m).
\end{split}
\end{equation}
Similarly $ \eta_t(m) \ge e^{-S(m)t} \bar{\eta}_t(m) $.
The result follows from the special case of the Palm integral
formula known as Campbell's Theorem \cite[(6.4.11)]{DV88}.
\end{proof}

\section{
Asymptotic Behavior}
\label{sec:asymptotic}
In contrast to the additive case, genotypes subject to subadditive
cost functions may tend to explode.  In age-structured models,
the total effect of mutant alleles acting after some given age is
limited, regardless of how many of them may accumulate.
If the rate at which some class of mutant alleles is generated
exceeds any countervailing selection-cost increment which
they may incur, the number of mutant alleles in that class
may be expected to grow without limit.

\begin{Thm} \label{T:blowup}
Assume the conditions of Theorem \ref{T:epistatic} are satisfied.
Let $B \subset\eu{M}$ be a subset with finite $\nu$-mass.
Suppose $ 0  \le S( g + g^*) - S(g) \le s $ for all $g$ and for
all those $g^* $ with masses only at points in $B$, that is,
with $ g^*(B) = g^*(\eu{M})$.  Let $ J_n^{*} $ be the indicator
function of the set of genotypes with $ g(B) = n $.
Then  $ s < \nu(B) $ implies that $ P_t J_n^* $ goes to zero for
every $n$ as $t$ goes to infinity.
\end{Thm}

\begin{proof}
We write our Poisson process $X_{t}$ as $X_{t}^{*}+X_{t}^{r}$,
where $X_{t}^{*}$ is the restriction of $X_{t}$ to $ B$
and $X_{t}^{r}$ is the remainder.  These components are
independent of each other.  Let $U :=\inf\{u: X_t(B) > 0 \}$
be the arrival time of the first point in $B$,  an exponential
random variable with mean $ 1/\nu(B) $.   We have
\begin{equation}
0 \le \int_{0}^{t} S(X^{r}_{u} + X^{*}_{u}) - S(X^{r}_{u}) du
  \le s(t - U ) \wedge 0.
\end{equation}
To bound $P_t J_n^*$ , we write the numerator of \eqref{E:epissol},
$ \tilde P_t J_n^* $,
as the expectation of a product of three factors,
$ \exp\bigl(-\int_{0}^{t}S(X^{r}_{u})du\bigr) $,
$ \exp\bigl(-\int_{0}^{t}S(X_u) - S(X^{r}_{u})du\bigr) $
and $ J_n^*(X^{*}_t) $.  The second factor is bounded above
by $1$ and the third factor is independent of the
first.  The denominator of \eqref{E:epissol},
$ \tilde P_t I $, has the same first factor,  the same second
factor,  bounded below by $ \exp ( -s(t-U) \wedge 0 ) $,
and a third factor identically equal to $1$.  Using
independence, we may cancel the expectations of the first
factors in numerator and denominator, so that $ P_t J_n^{*} $
is less than or equal to the quotient of
$ \mathbb{E} J_n^*(X^*_t) $
and $ \mathbb{E} \exp (-s(t-U) \wedge 0 ) $.
Writing $ \nu^* $ for $ \nu(B) $, this quotient equals
$(\nu ^* t)^{n}/n! $  divided by
$ ( \nu ^* e^{(\nu ^* -s)t}-s ) / ( \nu ^* - s ) $.
The quotient goes to $0$ as $t\to\infty$ for every $n$.
\end{proof}


\section{
Non-epistatic cost functions}
\label{sec:nonepistatic}

In the non-epistatic case, when the cost function $S$ is
additive,  a proof of uniqueness and an eminently
computable formula can be obtained which lead to conditions
for convergence as $t$ goes to infinity:


\begin{Thm} \label{T:additive}
Suppose that $S$ is an additive (nonepistatic) cost function
such that the expectation value $ \nu (S \wedge 1 )$ is
finite and suppose that $P_{0}$ is an initial probability measure
such that $P_{0}S$ is finite.
Then the equations \eqref{E:evfunc} have a unique solution
on $[0,\infty)$.  A random measure chosen according
to $P_{t}$ may be represented as the sum of two independent
random measures. The first component is a Poisson random measure with
intensity $ (1/S(m)) ( 1 - e^{-S(m)t})  d\nu(m). $
The second is the initial measure $P_{0}$, tilted by
the weighting $ e^{-tgS} $. That is,
the second component $Q_{t}$
satisfies $ Q_{t} F = \tilde Q_t F / \tilde Q_t 1 $ with
\begin{equation}
\tilde Q_t F = \int e^{ -S(g)t } F(g) dP_{0}(g)
\end{equation}
If $\nu$ is finite, this solution is identical with that given in
Theorem \ref{T:epistatic}.
\end{Thm}

\begin{proof}
Linearity of $S$ allows us to transform Equation \eqref{E:evfunc}
into a first-order linear partial differential equation.
Suppose we are given an integrable  non-negative
function $\sigma(t)$  which serves as a candidate for $P_t S$.
Let $z$ be a positive real number and let $f$ on $\eu{G}$
be a bounded nonnegative function with bounded support.
We take our test functions $F$ now to be of the
combined form $ F(g) = e^{-gf-zS(g)} $.
We write $ h(t,z) $ for the real function which will turn out
to be $ \log(P_t F )$ satisfying given boundary conditions
\begin{equation}
h(0,z) = \eta(z) = \log P_0 F.
\end{equation}

Thanks to the form of $F$ and the linearity of $S$,
the expression $ \int (F(g+\delta_m)-F(g)) d\nu(m) $
from \eqref{E:evfunc}  now equals $ F(g) \zeta(z) $, where
\begin{equation}\label{E:zeta}
\zeta(z) := \nu ( F(\delta_m) - 1).
\end{equation}
Since $f$ is non-negative,  $ \lv \zeta(z)\rv $
is bounded by $\nu(f)+(1+z)\nu( S \wedge 1) $.
The first term is finite because $f$ is bounded
with bounded support.  The second term is finite by
assumption.  If $ \exp(h) $ is to satisfy \eqref{E:evfunc},
we need $ h $ to satisfy the following partial differential equation
of the McKendrick type familiar to demographers:

\begin{equation}  \label{E:htz}
 \frac{\partial h(t,z)}{\partial t} -
 \frac{\partial h(t,z)}{\partial z} = \zeta(z) + \sigma(t).
\end{equation}
We have shown that the right-hand side is
well-defined for all non-negative $z$ and $t$.

We solve \eqref{E:htz} uniquely for $h$ by exploiting 
the method of characteristic curves to tranform it into 
a system of ordinary differential equations.
The characteristic curve passing through the
point $(t,z)$ is the line $\tau \mapsto (\tau,t+z-\tau)$
\cite[Section 3.2] {lE98}.
Defining $ \tilde{h}(\tau):=h(\tau,t+z-\tau)$,
we get $ \tilde{h}'(\tau)=\gs(\tau)+\zeta(t+z-\tau) $
for  $ 0\le \tau\le t+z$.
Integrating this equation from $0$ to $t$ gives
\begin{equation}  \label{E:hsol}
h(t,z) = \eta(t+z)+\int_{0}^{t} \gs(\tau)d\tau
          +\int_{z}^{t+z} \zeta(r) dr
\end{equation}
The final term in $\zeta$ is equal to
\begin{equation}  \label{E:hsolzeta}
 \nu \left[ -t +  \left( e^{-f(m)-zs(m)}- e^{-f(m)-(z+t)s(m)}
                           \right)/s(m) \right].
\end{equation}

We now set $ P_t F = \exp( h(t,z))$.
Additivity of $S$ makes the derivative of $P_t F$
equal the right-hand side of \eqref{E:evfunc}
plus $  ( \sigma(t) - P_t S )(P_t F) $.
Also, $-P_t S $ is the partial derivative of $h$ with
respect to $z$ at $z=0 $ and $ f \equiv 0 $, which is the sum
of $ \nu ( 1 - e^{tS}) $ and $ P_0 S e^{-tS}/P_0 e^{-tS} $.
Setting $-\sigma(t)$ equal to this sum is therefore the unique
choice which makes $P_t$ satisfy \eqref{E:evfunc}.
Writing out $h$ and setting $z = 0$, we recognize the
Laplace functional of the convolution of probability measures
specified in the theorem.
\end{proof}

The first piece of $P_{t}$ clearly converges to a Poisson random
measure as long as $ \nu/S $ is boundedly finite.
But that is not the complete story of asymptotic behavior.
In general, the influence of $P_0$ in $Q_{t}$ may persist.
In the limit, however, we may apply Varadhan's
Lemma \cite[Theorem III.13]{fdH00} to show that $Q_{t}$ becomes
concentrated on the set of genotypes of minimum selective cost.

\begin{Cor}  \label{C:wholeconv}
Suppose $P_{0}$ and $S$ satisfy the conditions of Theorem
\ref{T:additive},  that the support $\on{supp}\, P_{0}$ 
is compact, and that $S$ is continuous.
Let $\gs=\inf\{S(g)\, : \, g\in\on{supp} P_{0}\}$.
If $\eu{O}$ is any open neighborhood
of $\{g\in \on{supp} P_{0}\, : \, S(g)=\gs\}$, then
$ \lim_{t \to \infty} Q_{t}(\eu{O})=1$.
In particular, if $P_{0}\{0\}>0$, the tilted
measure $Q_{t}$ converges to $\delta_{0}$.
\end{Cor}

%


\section{
Applications to the theory of longevity}
\label{sec:longevity}
We outline a few of the many applications to the
biodemography of longevity.
We take the space of potential mutations $\eu{M}$  to be $C[0, \infty)$,
the continuous real-valued functions on $\R^{+}$, supplied
with any of the usual metrics corresponding to uniform
convergence on bounded intervals \cite[Section 1.44]{wR91}.
A mutation measure $\nu$ on this space is the distribution
of a stochastic process.   
We base our selective cost function $S(g)$
on Equation 4.9 of Charlesworth \cite[p. 140]{bC94},  
taking into account more recent discussion \cite[p. 930]{bC00a}.  
The cost function is defined in terms of the 
age-specific survival function $l_x(g)$ and the age-specific
fertility rate $f_x(g)$ specific to each genotype, along 
with a conversion factor $T$, representing a baseline value 
for the length of a generation, and a rate $r_0$ 
representing a population-wide baseline intrinsic rate
of natural increase usually set to zero in applications.  
\begin{equation}  \label{E:nrrcost}
S(g) =  ( 1 - \int e^{-r_0 x} l_x(g) f_x(g) dx )(1/T) 
\end{equation}


\subsection{
Gompertz hazards}\label{sec:gomp}
Charlesworth \cite{bC01} has suggested a possible
origin for Gompertzian (exponentially increasing)
hazard rates through a process of mutation-selection
balance which fits into our generalized model.
Members of a species are taken to be subject to a
common high background age-independent hazard
rate $ \lambda $ plus age-dependent contributions from
mutations.  Each mutant allele $m$ may be represented as a
continuous function $m(x)$ of age added onto the
hazard function for an individual.
Charlesworth's elementary models assume constant
fertility at all ages above an age $b$ of sexual
maturity, forgoing any
\it
a priori
\rm
upper age cutoff.

The selective cost $ S(g) $ for a genotype $g$ 
from \eqref{E:nrrcost} takes the following form when
time is measured in generations rather than years: 
\begin{equation}\label{E:fullcost}
S(g)  =   1 - \int_b^{\infty} \lambda \exp \Bigl(-\lambda x
      + \lambda b - \int \int_0^x m(a) da \, dg(m) \Bigr) dx  
\end{equation}
This cost function $ S $ is a non-additive epistatic cost function.
Following established practice, 
Charlesworth substitutes the additive cost function
\begin{equation}\label{E:addcost}
\hat{S}(g) =   \int \  \Bigl( \int_0^{\infty}  ( e^{-\lambda (x-b)} \wedge 1 )
                      m(x) dx  \Bigr) \  dg(m).
\end{equation}
This function $ \hat{S} $ is an additive approximation to $S$.  

We first show that under the same premises as \cite{bC01} 
our model confirms the same conclusions.   
With additive costs as in \eqref{E:addcost},
Theorem \ref{T:additive} and Corollary \ref{C:wholeconv}
give us sufficient conditions for the distribution of genotypes
to converge to the  Poisson random measure with
intensity $\nu/\hat{S} $.  It suffices that the starting
state put positive weight on the null state and have compact
support and that $\nu$ and $\nu/S$ be boundedly finite.
Then the average of the hazard rates over genotypes will converge
to $ \lambda  + \int_{\eu{M}} (m(x)/\hat{S}(m)) d \nu(m) $,
equivalent to \cite[Eq.4a]{bC01}.    

It is worth mentioning that this expression for the average
of the hazard rates is not the equilibrium aggregate hazard 
rate for the whole population,  because the heterogeneity 
mediated by the Poisson distribution implies attrition of 
higher-risk genotypes with advancing age.  The Poisson expression
for the additive genetic variance and covariance also
require modification for age-specific attrition.

Charlesworth focuses on translation families of mutations,
which we may write as $m_y(x) = m_0(x-y) $ with effects only
after an age of onset $y$.  With $ d\nu(m_y) = \nu_0 dy $
on some $ [b^{\prime}, \infty) $, he displays choices for $m_0$
which make the average of the hazard rates into an exact Gompertz-Makeham
function  $ \lambda + \nu_0 \exp( -\lambda (x - b) ) $ on the
support of $\nu$,  and others which approximate Gompertz-Makeham
shapes for large $x$.    (These shapes do not include heterogeneity
corrections.)  

We now observe that our generalized model predicts different
qualitative behavior when the additive approximation 
of \eqref{E:fullcost}  by \eqref{E:addcost} is not guaranteed to hold.
The additive theory predicts that a ``wall of death'' 
with an infinite equilibrium mean hazard rate  
appears at, but not before, the age at which reproduction 
comes to an end \cite[p. 60]{bC01}. 
Theorem \ref{T:blowup} implies a more dramatic breakdown.
The mean hazard function can actually reach infinity
at ages at which fertility is still strictly positive,
if the full epistatic cost function $S$ in \eqref{E:fullcost}
is kept in place of the additive approximation \eqref{E:addcost}.
The same is also true, if the bounded cost function $S$ is
replaced by an unbounded cost function defined, as in 
Equation 4.12 of \cite[p. 141]{bC94},  
to equal the decrement to the intrinsic rate of natural increase
resulting from the mutations contributing to each genotype.  
Contrary to additive theory, the ``wall of death'' is not
tied to the end of reproduction but involves a fine-tuned
balance between mutation rates and tapering costs.

%
%

%

\subsection{
Gaussian process mutations}  
\label{sec:gauss}
We now apply our model to move beyond stylized cases and
investigate a wider range of possible specifications 
for the age-specific effects of mutations.
The cases considered in \ref{sec:gomp},  in which a constant
background mortality imprints a Gompertzian pattern onto
increments to the hazard function,  share the property
that every mutant is deleterious at every affected age.
Is this property essential to the imprinting,  or
can the age-specific force of selection readily produce the
same kind of outcomes with mutants that mix positive
and negative effects?

Our framework allows quite general pleiotropic
specifications.  A natural starting point is the case
of Gaussian processes.  The fitness cost for this brief
discussion will be the additive approximation \eqref{E:addcost}.
Suppose that the mutation process generates mutations proportionately
to a positive-real-parameter Gaussian process with expectation $a(x)$
and covariance function $c(x,x')$, conditioned on fitness cost
bounded away from 0.
That is, if we look at the pattern of age effects in a randomly
chosen mutation, it looks like a realization of this Gaussian
process, subject to $ \hat{S}(m) > s >0 $ for some $s$.
The overall rate of mutation is a constant $\nu_0$.
For rigorous treatment, we also need to condition on events
which keep the resulting hazard functions non-negative and
insure the validity of the additive approximation, but
here we shall assume that the choices of parameters
keep misbehavior rare enough that it can be neglected.


The average over genotypes of the hazard function is given by
\begin{equation}  \label{E:genmort}
h(x) = \lambda + \int_{\mM} \frac{m(x)d\nu(m)}
     {\int_{0}^{\infty}\gl e^{-\gl z} dz\int_{0}^{z+b} m(y)dy}.
\end{equation}
The denominator $ \hat{S}(m) $ is obtained from \eqref{E:addcost}
by integration by parts.   Since the numerator and denominator
of \eqref{E:genmort}, linear functionals of $m$,
are both Gaussian random variables,
we can describe their joint distribution simply by computing their
covariance.  The conditional expectation of $m(x)/ \hat{S}(m) $
conditional on $ \hat{X}(m) $, is obtained via linear regression.


When $a(x)\equiv 0$, the conditional expectation turns out
to be independent of $\hat{S}(m)$, making the integral
in \eqref{E:genmort} independent of the bound $s$ (except
for a small change in the proportionality constant) and equal
to $ \nu_0 \, \mathbb{E} \, m(x) \hat{S}(m) /\on{Var}\lp S(m)\rp.$
For processes with zero mean,
then, we may treat $s$ as 0.
As one example, take the mutations to be realizations of
Brownian motion, so that $c(x,x')=x\wedge x'$ and $a(x)=0$.
The increment to the mean hazard rate then has the
form $c_{1}-c_{2}e^{-\gl x}$ for $x\ge b$.

Can any Gaussian mutation process with zero mean generate
approximations to Gompertz-Makeham hazard functions?    The
covariance kernel must satisfy $\lv c(x,y)\rv\le c(x,x)/2 +c(y,y)/2$.
With $a(x)\equiv 0$, the incremental mortality is bounded
above by $c_{1}+c_{2}c(x,x)$
for constants $c_{1}$ and $c_{2}$.  The mortality thus cannot be
exponentially increasing over a long range of ages, unless this exponential increase is built into the mutation process itself.


\section{
Historical Background}
\label{sec:prior}
We discuss in this section the relationship of our model
to the existing corpus of work on related topics.
It was J.B.S. Haldane \cite{jH37} who articulated the concept of
mutation-selection balance as early as 1937.
Crow and Kimura \cite{CK70}, Ewens \cite{wE79}, and
Kingman \cite{jK80} give the foundations of the subject.
B\"urger \cite{rBu98} and \cite{rB00} covers the present
state of the art.  These authors put only limited emphasis
on age structure;  Charlesworth \cite{bC94}
and \cite{bC00} propounds the age-specific side.

Infinite population models in which fitness is a function
of the number but not the identity of mutant genes
go back to Kimura and Maruyama \cite{KM66}.
They state discrete-time and continuous-time dynamic
equations for special cases which readily suggest
the general ``mutation-counting model''
\eqref{E:timeevolsimp}. They obtain
some closed-form equilibrium distributions.
Conditions for convergence to stationary states follow
from a theorem of Kingman \cite{jK77}, generalizing
theorems of Moran \cite{pM76,pM77}.
B\"{u}rger \cite[pp. 298-308]{rB00} traces the subsequent history.
Markov-chain versions with stepwise mutations of
identical deleterious effect have Poisson stationary
distributions (see e.g. Haigh \cite{jH78}
or Durrett \cite[p. 137] {rD02}).
The Poisson limit is implicit in estimations of
equilibrium genetic variance \cite{bC01}.

Mutation-counting models in the tradition of Kimura and
Maruyama are more tractable than the general case
considered here because they are a kind of
multi-locus model that can be subsumed under the
theory of single-locus models.  The count of mutant
alleles at different sites can be likened to the
integer label on a countable set of alleles
at a single site, subject to constraints on the
non-zero interallelic mutation rates.  Models defined
by various alternative sets of constraints have been
studied in some detail.

The most famous of these single-locus models is
Sir John Kingman's ``House of Cards'' (HC)
described in \cite{jK78} and \cite{jK80}.
Kingman's infinite-population discrete-time model posits
a single gene with potentially infinitely many alleles.
Alleles mutate to new alleles at a constant rate;
each new allele has a random fitness, given by a
probability distribution on $[0,1]$.  The state of the system
is given by a distribution of fitnesses on $[0,1]$, and the
dynamics are governed by a standard evolution equation.
Kingman \cite{jK78} gives the original proof that the distribution
of fitnesses for the HC process converges to a limiting distribution.
This model has many descendants, including the
``HC-approximation'' \cite{mT84} for stabilizing selection
around an intermediate optimum.

Our model differs from HC and its counterparts in four main ways.
Mutant alleles in HC have no properties other than fitness.
Mutant alleles in our model are tagged by an effect
represented by a point in a general metric space
whose specification determines the fitness through the
impact on demographic rates.
In HC there is only a single locus.  In our model we are
concerned with the heterogeneity of whole sets of mutant alleles
across a large number of loci within the population.
Because HC includes only a single locus, it offers no possibility
for interactions between the fitnesses of different alleles.  Our
model is open to general epistasis.
Finally, HC is well-suited to the use of Markovian methods and
sample-path analysis,  whereas our proofs require non-Markovian
machinery.

A highly versatile general formulation of single-locus models
has been developed by Reinhard B\"urger \cite[chapter IV.2]{rB00}.
His ``general mutation-selection model''
warrants close comparison with our own.
Like us, B\"urger draws mutants from a general space,
according to an arbitrary distribution.
B\"urger requires his space to be locally compact, whereas we allow
any complete, separable metric space so as to include
mutants identified with continuous functions on $[0,\infty)$.

The substantive difference between B\"urger's model and our own is
in their contrasting views of the genome.
B\"urger focuses on a single locus,
with (perhaps) infinitely many potential alleles.  Each individual's
genotype is characterized by a single quantity and the population
is characterized by a distribution on the mutation space.
We take a more synoptic view, watching the (perhaps)
infinitely many alleles pop up at (perhaps) infinitely many loci,
thereby opening up the representation of population
heterogeneity.  In our model, the only mutation
process is the conversion of an undifferentiated wild type
to a random mutant allele, so we need not introduce transition
rates between alleles. However, our flexible treatment of
heterogeneity means that even the description of the state of the
system has to be more abstract than is customary in population
genetics.

In a different setting, Del Moral and Miclo \cite{dM00}
present results which parallel our Theorem \ref{T:epistatic}.
Their conditions are more general in some respects and more
restrictive in others.  While their concerns are remote from
biology, they use the terminology of ``mutation generators''
and ``adaptation''  in their descriptions.  They prove that
the differential equation model which we analyze can be
derived as an infinite-population limit of finite nondeterministic
Moran models for interacting particles.
A  new book \cite{dM04} expands this line of investigation.

We accompany Del Moral and Miclo, on a road that diverges from
the Markov modeling, branching
processes, branching diffusions, and superprocesses which are so important
to stochastic population theory \cite{EK86}.  Pioneering work in these areas
by \cite{sS76,rG79,rG83} has been followed by extensive results
on particle processes and measure-valued diffusions with selection,
including \cite{DK99,EK87,EK93,DF98,DF01,KKN95, MR2003k:60104, MR94m:60101}.
Such Markov processes, even when they involve
selection, are essentially  linear.  Lineages rise or fall at their
own rates, according  to their fitnesses, independent of the outside
population.  By contrast, the mutation-selection paradigm on which
we focus has to be nonlinear, since every  lineage has negative fitness.
The models are saved from trivial degeneracy by a renormalization,
conditioning the process on long-term survival.
This ingredient introduces a  quadratic nonlinearity into the evolution
equation, inasmuch as the entire  population contributes
to the selective pressure on each individual, bringing
non-Markovian arguments to the fore.

\section{
Prospects} \label{sec:prospects}

The generalized model for mutation-selection balance presented
here can be applied widely to settings where age structure matters.
Because the model allows mutations with a mixture of positive
and negative effects,  it gives scope to some blending of ideas
about mutation accumulation with ideas about antagonistic pleiotropy.
It offers a handle on responses to changing fitness
conditions through the finite-time solutions,
along with machinery for treating epistatic cost functions.

The Palm formula in Section \ref{sec:represent}
facilitates the construction of an alternative
version of our model which, in contrast to
\eqref{E:evfunc},  allows for free recombination (FR).
In line with \cite{KM66}, \cite{pH94}, and \cite{kD99},
we postulate conditions on the relative rates of
recombination, selection, and mutation which lead,
in the continuous-time limit, to a process in which
$P_t$ is always a Poisson random measure on $\eu{G}$
with some intensity measure $\rho_t$.
Our results in Section \ref{sec:represent}, derived in the
absence of recombination, give us the form of an equation
for $\rho_t$ in this generalized free-recombination model.
Differentiating \eqref{E:palm} at $ t = 0 $ leads
to a representation for $ \rho_t $ of the form
\begin{equation} \label{E:FR}
\rho_t = \rho_0 + \nu t - \int_0^t D \rho_{\tau} d\tau
\end{equation}
Here $ D \rho $ is a measure whose density with respect to $\rho$
at $m$ is $ \mathbb{E}[S(X^{\rho} + \delta_m) - S(X^{\rho})] $, and
$ X^{\rho}$ is the Poisson random measure with intensity $\rho$.
Rigorous development of this alternative is reserved
for a future paper.




\section*{Acknowledgements}
D.R.S. was supported by Grant K12-AG00981 from the National
Institute on Aging,  S.N.E. was supported in part by
Grants DMS-0071468  and DMS-0405778 from the National Science Foundation and by the
Miller Foundation, and  K.W.W. was supported by Grant P01-008454
from the National Institute on Aging.  The authors thank Montgomery
Slatkin and Nick Barton for helpful discussions.

\end{document}